\documentclass[11pt, a4paper]{article}
\pdfoutput=1

\usepackage{amsmath}
\usepackage{amsfonts}
\usepackage{amssymb}
\usepackage{graphicx, rotating}
\usepackage{epstopdf}
\usepackage{mathrsfs}
\usepackage{epsfig}
\usepackage{latexsym}
\usepackage{graphicx}
\usepackage{color}
\usepackage{amsmath,amssymb}
\usepackage{cite}
\usepackage{slashed}
\usepackage{hyperref}
\usepackage{datetime}

\numberwithin{equation}{section}

\hypersetup{colorlinks, citecolor=bluscuro, linkcolor=black, urlcolor=bluscuro}
\definecolor{rossos}{rgb}{0.8,0.2,0.3}
\definecolor{bluscuro}{rgb}{0.15, 0.2, .85}
\definecolor{bluchiaro}{cmyk}{1,.3,0.,0.1}

\makeatother   
\newcommand{\GeV}{{\rm \,GeV}}

\newcommand{\cm}{{\rm \,cm}}

\def\de{\textrm{d}}

 \def\be   {\begin{equation}}   \def\ee   {\end{equation}}
 \def\ba   {\begin{array}}      \def\ea   {\end{array}}
 \def\bea  {\begin{eqnarray}}   \def\eea  {\end{eqnarray}}
 \def\bean {\begin{eqnarray*}}  \def\eean {\end{eqnarray*}}
 \def\nn{\nonumber}

\begin{document}


%
\begin{flushright} 
CERN-PH-TH/2013-091\\
SISSA 19/2013/FISI
\end{flushright}

\vspace{1.5cm}
\begin{center}

{\huge \textbf {
On the Minimum Dark Matter Mass \\
[0.4cm]
Testable by Neutrinos from the Sun
}}
\\ [1.5cm]
{\large{\textsc{
Giorgio Busoni$^{\,a,\,}$\footnote{giorgio.busoni@sissa.it},  
Andrea De Simone$^{\,a, b,\,}$\footnote{andrea.desimone@sissa.it},   
Wei-Chih Huang$^{\,a,\,}$\footnote{wei-chih.huang@sissa.it}  
}}}
\\[1cm]

\large{\textit{
$^{a}$~SISSA and INFN, Sezione di Trieste, via Bonomea 265, I-34136 Trieste, Italy
\\  \vspace{1.5mm}
$^{b}$~\textit{CERN, Theory Division, CH-1211 Geneva 23, Switzerland}
}}
\end{center}

\vspace{1cm}

\begin{center}
{\large
\textbf{Abstract}
\begin{quote}
We discuss a limitation 
on extracting bounds on the scattering cross section
of dark matter with nucleons, using neutrinos from the Sun.
If the dark matter particle is sufficiently light (less than about 4 GeV), the effect
of evaporation is not negligible and the capture process goes in equilibrium with the 
evaporation.
In this regime, the flux of solar neutrinos of dark matter origin
becomes independent of the scattering cross section and therefore no constraint
can be placed on it.
We find the minimum values of
dark matter masses for which the scattering cross section on nucleons
can be probed using neutrinos from the Sun.
We also provide simple and accurate fitting functions
 for all the relevant processes of GeV-scale dark matter in the Sun.
\end{quote}
}
\end{center}

\def\thefootnote{\arabic{footnote}}
\setcounter{footnote}{0}
\pagestyle{empty}

\newpage
\pagestyle{plain}
\setcounter{page}{1}

\section{Introduction}

The indirect searches for the Dark Matter (DM) component of the Universe
are primarily based on identifying excesses in fluxes of cosmic rays, such as 
positrons, anti-protons, neutrinos, etc;
these  stable Standard Model particles may be the end product of 
 the annihilation (or decay) of DM in the galactic halo or in the Sun.
Among the different ongoing search strategies, the search for the annihilation
products of DM in the Sun is particularly interesting.
In fact, the DM particles trapped in the core of the Sun may annihilate into anything, but only
neutrinos would be able to escape the surface and reach the Earth.
The role of neutrinos in DM searches of this type is then very special.

The indirect searches for DM in the Sun are tightly linked to direct detection searches,
which are sensitive to the cross section for DM scattering off the nucleons of heavy nuclei (e.g. $\sigma_p$ for protons).
In fact, suppose that DM annihilates into several final states $j$, with branching ratios BR$_j$,
and producing a differential number of neutrinos per annihilation $\de N_j/\de E_\nu$;
then, the flux of neutrinos of DM origin arriving at Earth is given by
\be
\frac{\de\Phi_\nu}{\de E_\nu}=\frac{\Gamma_A}{4\pi R^2}
\sum_j \textrm{BR}_j \frac{\de N_j}{\de E_\nu}
\ee
where $R$ is the Sun-Earth distance, $\Gamma_A$ is the rate of annihilations per unit time
\be
\Gamma_A=\frac{1}{2}A_\odot N_\chi^2\,,
\label{GammaA}
\ee
$N_\chi$ is the number of DM particles in the Sun, and the annihilation coefficient $A_\odot$
will be defined and discussed later, see Eq.~(\ref{Arate}).
Since $N_\chi$ depends on how many DM particles got trapped in the Sun, and hence generically
depends on $\sigma_p$, observational limits on the flux $\Phi_\nu$ translate into limits on $\sigma_p$, which can be competitive with those of direct detection searches.

This situation  dramatically changes in the case of light DM, with mass around GeV.
The number $N_\chi$  becomes independent
 of $\sigma_p$ when the capture process goes in equilibrium with the evaporation,
 and annihilation is negligible. As a consequence, the experimental bounds on the neutrino
 flux from the Sun cannot be translated anymore into constraints on $\sigma_p$ and the link between neutrino flux and DM-nucleon scattering cross section disappears.
We also find  simple and accurate fitting functions for all the relevant processes concerning
DM in the Sun: annihilation, capture and evaporation.

The interest in $\mathcal{O}$(GeV) neutrinos as probes of DM 
has been recently reinvigorated  by the proposal to consider the production
 in the Sun of muons and charged pions as products of DM annihilations, and
 their subsequent
decay at rest \cite{rott, Bernal:2012qh}.
These neutrinos can be easily detected by neutrino telescopes
based on water Cherenkov detectors, such as  \textsc{Super-Kamiokande} \cite{superk}.
One should also keep in mind that 
the energy to distinguish neutrinos  originated by DM in the Sun is bounded from below; in fact, 
for  DM masses below $\sim 100$ MeV, the detection process is based
on inverse $\beta$-decay
$\bar\nu_e+p\to e^+ + n$, of which $e^+$ gets identified.
The distribution of $e^+$  is mostly isotropic (see e.g. Ref.~\cite{Vogel:1999zy}), and the angular resolution is typically not good enough to extract
information on the arrival direction. Therefore, it is not possible to distinguish neutrinos from DM annihilations in the
Sun from those from the galactic halo, whose flux is  much bigger \cite{Yuksel:2007ac}.
While we will not commit ourselves to any specific model for GeV-scale DM,
this situation can be realized in the context e.g. of asymmetric DM \cite{asymDM} or in
explicit models such as the one in Ref.~\cite{Allahverdi:2013mza}.

The rest of the paper is organized as follows.
In Section \ref{sec:processes} we will briefly discuss the relevant processes
for DM inside the Sun, and then turn to
compute the total number of the DM inside the Sun in Section \ref{sec:number}.
Our concluding remarks are in Section \ref{sec:conclusions}.

\section{Relevant processes of DM in the Sun}
\label{sec:processes}

The DM inside the Sun undergoes several processes: it gets
captured, via the energy losses from scattering with the nuclei; it annihilates, whenever two
DM particles meet; or it can even evaporate, if the collisions with nuclei make it escape the Sun.
The total number of DM inside the Sun is thus determined by the interplay of these three processes.
Let us discuss them in more detail (see also Ref.~\cite{Jungman:1995df} for a previous
analysis of these processes, and Ref.~\cite{Kumar:2012uh} for a recent update in the regime where evaporation is not important).

\subsection{Annihilation}
The first important process to consider is the annihilation of two DM particles inside the Sun,
and we want to compute the rate for this process (we follow closely the discussion in Ref.~\cite{griest}).
We approximate the phase space distribution of the DM trapped in the Sun by  a global temperature $T_\chi$ and the
local gravitational potential $\phi(r)$, defined with respect to the solar core, as
\be
\phi(r)=\int_0^r {G_N M_\odot(r')\over r'^2} \de r'\, ,
\label{phi}
\ee
where $G_N$ is Newton's constant and $M_\odot(r)=4\pi\int_0^r r'^2 \rho_\odot(r')\de r'$ is the solar mass within radius $r$.
Throughout the paper we use the density profile $\rho_\odot(r)$  from the  
solar model AGSS09 \cite{solar}. 
The DM number density  is determined by solar gravitational potential 
and scales as
\be
n_\chi(r)=n_0 e^{-m_\chi \phi(r)/T_\chi}\,,
\ee
where $n_0$ is the density at the core. 
The annihilation coefficient $A_\odot$ in Eq.~(\ref{GammaA}) is defined as
\be
A_\odot\equiv\langle \sigma v_{rel} \rangle_{\odot}\frac{\int_{\textrm{Sun}} n_\chi(r)^2 \, \de^3 r}{\left[\int_{\textrm{Sun}} n_\chi(r) \, \de^3 r\right]^2}\,,
\label{Arate}
\ee
where 
the thermally-averaged annihilation cross section $\langle \sigma v_{rel} \rangle_{\odot}$ is assumed
to be independent on the DM position in the Sun,
 and we assume the number density of DM particles  equal to that of antiparticles.
 The factor of 1/2 in Eq.~(\ref{GammaA}) simply avoids double counting of pairs in the annihilation.

\begin{figure}[t]
\centering
\includegraphics[width=7.5cm,height=5cm]{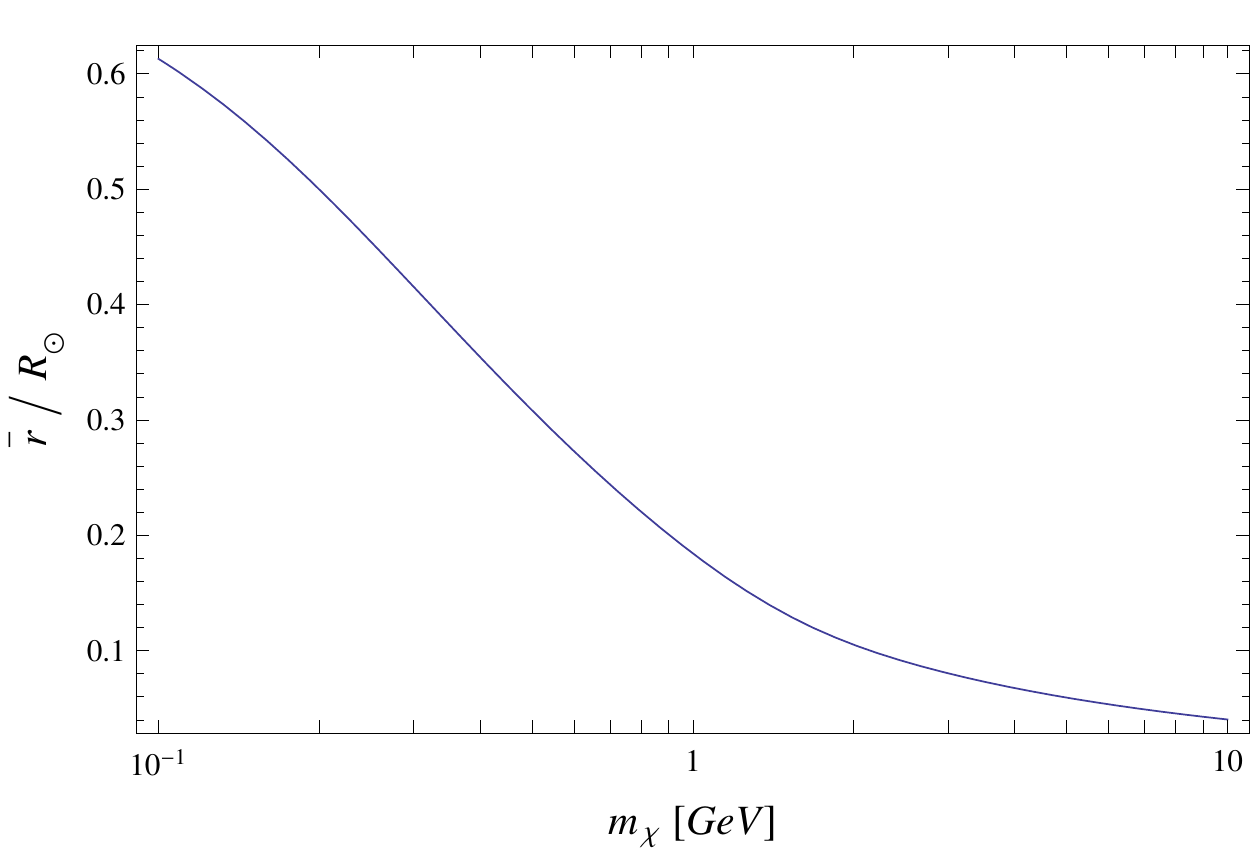}
\hspace{0.3cm}
\includegraphics[width=7.5cm,height=5cm]{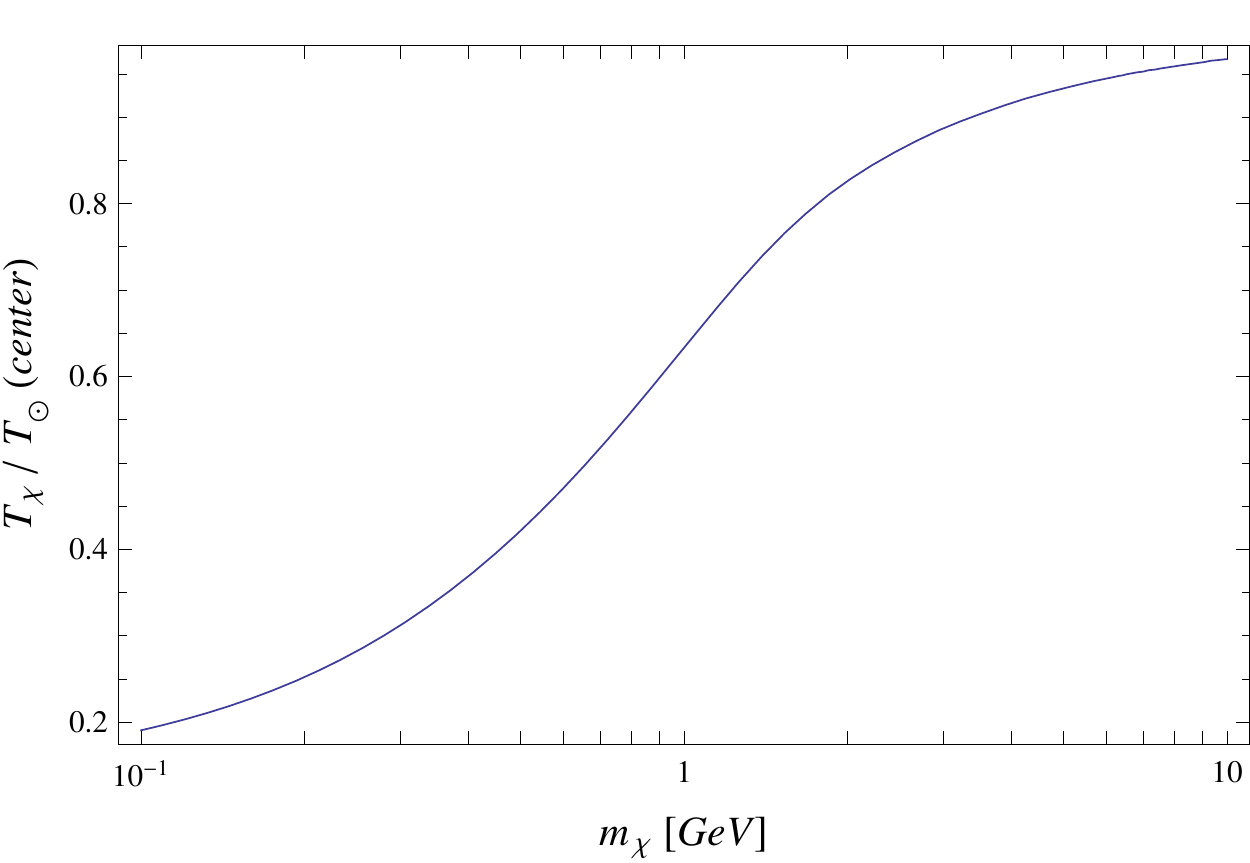}
\caption{\emph{{\small
\emph{Left panel:} The average orbit radius $\bar r$, normalized to the solar radius,
as a function of the DM mass.
\emph{Right panel:}
The temperature of the population of DM particles trapped in the Sun $T_\chi$, normalized to the central solar temperature, as a function of the DM mass.
}}
}
\label{fig:TofM}
\end{figure}

To compute the annihilation coefficient $A_\odot$, we need to know $T_\chi$, which is obtained as follows.
 The average DM orbit radius $\bar r$ is the mean value of the DM distance from the
center of the Sun,
\be
\bar r(m_\chi)={\int_{\textrm{Sun}} r\, n_\chi(r) \de^3 r\over
\int_{\textrm{Sun}} n_\chi(r) \de^3 r}\,,
\ee
and it depends on the DM mass (see Fig.~\ref{fig:TofM}, left panel).
The temperature of the population of DM particles trapped in the Sun,
or DM temperature $T_\chi$ for brevity, is taken to be the local solar temperature at the DM mean orbit:
\be
T_\chi=T_\odot(\bar r)\,,
\ee
and it depends on $m_\chi$ through $\bar r$.
The dependence of $T_\chi$ on the DM mass is shown in the right panel of Fig.~\ref{fig:TofM}.
If the DM particles are heavier than a few GeV, they get trapped near the solar core
and the corresponding $\bar r$ will be very small. As a consequence, the DM temperature
$T_\chi$ will be close to the central solar temperature. In the limit where the  DM is much heavier
than the nucleon mass $m_\chi \gg m_N$, the DM temperature will approach the solar temperature
at the center.
The determination of  $T_\chi$, and hence the annihilation coefficient, for  DM masses of a few GeV  (or less) requires taking into account
the full solar density profile, as the DM orbit can span a wide region inside the Sun and the approximation of constant solar density is no longer valid.

\begin{figure}[t]
\centering
\includegraphics[scale=1]{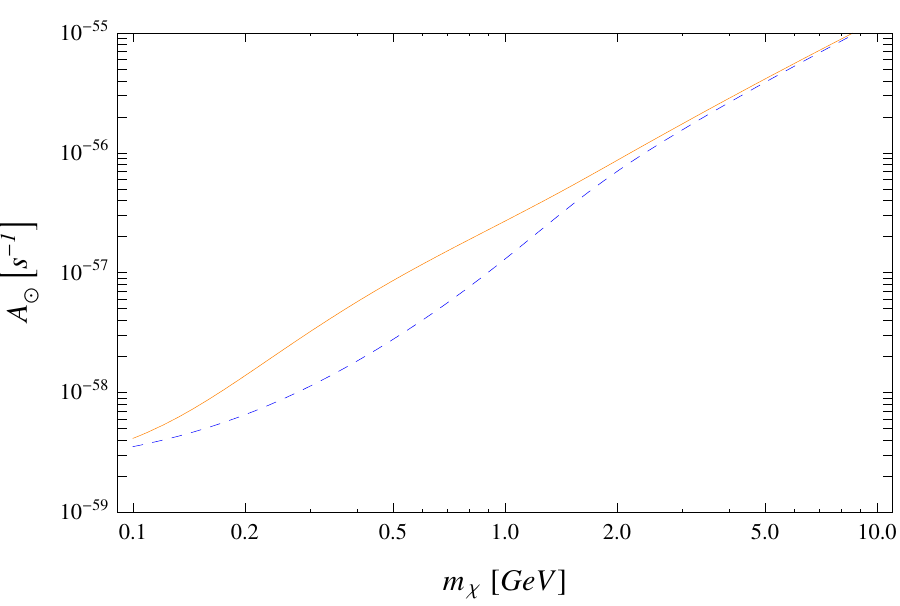}
\caption{\emph{{\small
The annihilation coefficient $A_\odot$ for $\langle \sigma v_{rel} \rangle_\odot=3\times 10^{-26}\, \rm{cm}^3/\rm{s}$ (orange solid line). We compare  with Eq.~(16) of Ref.~\cite{kw} (blue dashed line):
$A_\odot=
(\sqrt{2}/\pi\bar r)^3
 \langle \sigma_{{\rm ann}} v \rangle_{\odot}$.
}}}
\label{fig:Ann_plot}
\end{figure}

For the annihilation coefficient $A_\odot$  we find the following fitting function
\bea
A_\odot\simeq 2.91\,
e^{-1.34\left[ \log\left(\frac{20 \GeV}{m_{\chi}}\right) \right]^{1.14}}
 \left(\frac{ \langle \sigma_{\rm ann} v \rangle_{\odot} } {3\times 10^{-26} \rm{cm}^3/\rm{s}}\right) 
 \times 10^{-55} \,\rm{s}^{-1}\,,
\label{Aratefit}
\eea
valid in the range $0.1 \GeV \leq m_\chi\leq 10$ GeV, to better then 9\%.
 In Fig. \ref{fig:Ann_plot}, we show the comparison between our numerical results with Eqs.~(16) of Ref.~\cite{kw} for $ \langle\sigma_{\rm ann}v\rangle_\odot=3\times 10^{-26} \rm{cm}^3/\rm{s} $. They are consistent with each other up to $m_\chi=1$ TeV, except for $m_\chi \leq 2$ GeV, which is
 due to the breakdown of the constant density approximation.

In the following, we will only consider  the case where the annihilation cross section
is  velocity-independent ($s$-wave annihilations). 
As a reference value for the thermally average cross section in the Sun today we take 
$\langle\sigma_{\rm ann}v\rangle_\odot=3\times 10^{-26}\cm^3/\rm{s}$,
although the actual value depends on the effective degrees of freedom at the freeze-out temperature,
which in turn depends on the DM mass (see e.g.~Ref.~\cite{gondolo}).
The case of pure $p$-wave annihilations results in a smaller annihilation cross section today than at freeze-out.
We will not explore this case thoroughly, although in our analysis we will vary the annihilation cross section
with respect to its reference value.

\subsection{Capture}

The other relevant processes occurring in the Sun are capture and evaporation. 
A DM particle can collide with nuclei and lose energy when it traverses the Sun.
If the final velocity of the DM particle after the collision is less than the local escape velocity $v_e(r)$, then
it gets gravitationally trapped. This capture process makes the popoulation of DM particles in the Sun grow.
However,  the captured DM particles may scatter off energetic nuclei and be ejected, whenever the 
DM velocity after the collision is larger than the local escape velocity. This process is called evaporation.
The formalism to describe capture and evaporation is the same,
apart from the requirement on the final velocity to be larger or smaller than $v_e$.

The local escape velocity is defined as
$v_e(r)\equiv\sqrt{2\left[\phi(\infty)-\phi(r)\right]}$,
where $\phi(r)$ is the local gravitational potential in Eq.~(\ref{phi}).
The basic quantity is the rate per unit time $R_i^\pm(w\to v')$ at which a single DM particle of velocity $w$ scatters to a final velocity between $v'$ and $v'+\de v'$, off a thermal distribution of nuclei $i$ with number density $n_{N_i}$, mass $m_{N_i}$ and temperature $T_{N_i}=T_\odot(r)$. The plus (minus) sign refers to whether the final velocity is larger (smaller)
than the initial one.
This quantity has been first computed in Ref.~\cite{gould87}, under the assumption of isotropic, velocity-independent DM-nucleus cross section $\sigma_i$,
and we provide the details of the calculation  in Appendix \ref{app:A}.
The scattering rate per unit time  results to be (see Eqs.~(\ref{Rplus})-(\ref{Rminus}))
\be
R_i^\pm(w\to v')\de v'=\frac{\sigma_i n_{N_i}}{w}\frac{\mu_{+,i}^2}{\mu_i}
\left[\left[\textrm{Erf}(\alpha_{+,i})-\textrm{Erf}(\pm\alpha_{-,i})\right]
+ e^{ -\frac{m_\chi  ( v'^2 -w^2)}{2T_{N,i}}}
\left[ \textrm{Erf}(\beta_{+,i})-\textrm{Erf}(\pm\beta_{-,i})\right]
\right]v'\de v'
\label{Rplusminus}
\ee
where Erf$(x)$ is the error function and
\be
\begin{tabular}{rclrcl}
$\alpha_{\pm, i} $&$\equiv$& $\sqrt{\dfrac{m_{N_i}}{2 T_{N_i}} } \left(\mu_{+,i} v' \pm \mu_{-,i} w \right)\,, $&
$\beta_{\pm, i}$ &$\equiv$& $\sqrt{\dfrac{m_{N_i}}{2 T_{N_i}} } \left(\mu_{-,i} v' \pm \mu_{+,i} w \right)\,,$ \\
$\mu_i$&$\equiv$&$\dfrac{m_\chi}{m_{N_i}}\,,$ &$ \mu_{\pm, i}$&$\equiv$& $\dfrac{\mu_i \pm 1}{2}\,.$
\end{tabular}
\label{alphabetamu}
\ee
The  rate per unit time is simply related to the differential scattering cross section $\de \sigma_i$ by 
$R_i(w\to v')\de v'=n_{N_i} w \,\de \sigma_i$.
The rates per unit time $\Omega_{v_e,i}^{\pm}(w)$ are simply obtained by appropriate integrations
over the final DM velocity
\bea
\Omega^{-}_{v_e, i}(w)&=&\int_{\frac{|\mu_{-,i}|}{\mu_{+,i}}w}^{v_e}  R^-_i(w\to v')\de v'\,,
\label{omegaminus}\\
\Omega^{+}_{v_e, i}(w)&=& \int_{v_e}^{+\infty}  R_i^+(w\to v')\de v'\,.
\label{omegaplus}
\eea
The lower integration limit in Eq.~(\ref{omegaminus}) is the minimal final velocity simply  set by kinematics.
The rate $\Omega_{v_e,i}^-$ is what controls capture, while $\Omega_{v_e,i}^+$
controls evaporation. 
We discuss here the capture process  and defer  evaporation to the next subsection.

The local capture rate of DM per unit volume at radius $r$, due to nucleus $i$ of mass $m_{N_i}$,  
 can be written as  \cite{Gould:1987ir, Wikstrom:2009kw}
\be
\frac{d C_{\odot, i}}{dV}=\int_0^{u_i^{\rm max}} \de u \frac{f_{v_\odot}(u)}{u} w \Omega_{v_e,i}^{-}(w),
\label{dCdV}
\ee
where $u$ is the DM velocity at infinity, $w(r)=\sqrt{u^2+v_e(r)^2}$ is the local DM velocity inside the Sun 
before the scattering,  and
$u_i^{\rm max}\equiv v_e \sqrt{\mu_i}/|\mu_{-,i}|$ corresponds to a DM scattering with a final velocity
equal to $v_e$. 

The function $f_{v_\odot}(u)$ is the velocity distribution of DM particles seen by an observer moving at the velocity of the Sun $v_\odot\simeq 220$ km/s, with respect to the DM rest frame.
The velocity distribution of DM particles in the galactic halo, in their rest frame, is approximated by a Maxwell-Boltzmann $f_0(u)$
with a  velocity dispersion $v_d$
\be
f_0(u)=\frac{\rho_\chi}{m_\chi}{4\over \sqrt{\pi}}\left(\frac{3}{2}\right)^{3/2} \frac{u^2}{v_d^3} e^{-3 u^2/(2 v_d^2)}\,,
\label{MB}
\ee
where $\rho_\chi\simeq 0.3 \GeV/\cm^3$ is the average mass density of DM in the halo.
We will set $v_d=270$ km/s.
By making a Galilean transformation of velocity $v_\odot$, it is straightforward to derive the distribution $f_{v_\odot}(u)$  
\be
f_{v_\odot}(u)=\frac{\rho_\chi}{m_\chi} \sqrt{\frac{3}{2\pi}}\frac{u}{v_\odot v_d} 
\left[\exp\left(-\frac{3 (u-v_\odot)^2}{2 v_d^2}\right)-
\exp\left(-\frac{3 (u+v_\odot)^2}{2 v_d^2}\right)\right]\,.
\label{fvsun}
\ee
In the Sun, the solar temperature is much smaller than the escape energy $(1/2)m_\chi v_e^2$ 
of a DM particle,
so for capture it suffices to deal with the zero-temperature limit.
We checked that taking into account the finite-temperature
corrections would reduce  the capture rate by less than 10\% with respect to the one computed
for $T_{N_i}=0$.

In the limit $T_{N_i}=0$, and for elastic isospin-invariant contact interactions between DM and nuclei, 
simple analytical formulae can be derived.
The scattering rate per unit time for nucleus $i$ is
\be
R^-_i(w\to v')\de v'=2 \frac{n_{N_i} \sigma_i}{w}\frac{\mu_{+,i}^2}{\mu_i} v'\de v'\,.
\ee
and the total rate (\ref{omegaminus}) becomes 
\be
\Omega_{v_e, \textrm{H}}^{-}(w)=\frac{\sigma_\textrm{H} n_{N_\textrm{H}}}{w} \left( v_e^2 - \frac{\mu_{-,\textrm{H}}^2}{\mu_\textrm{H}} u^2 \right)\,,
\label{OmegaH}
\ee
which is valid only for Hydrogen (H). In fact, for scatterings with heavier elements
one should take into account the decoherence effect. One simple way to do so is 
to multiply the scattering rate $R$ by a form factor $|F_i(E_R)|^2$, depending on the recoil
energy, which is the difference between the energies of the DM particle before and after
the collision $E_R=(1/2)m_\chi (w^2-v'^2)$.
So for Hydrogen $|F_\textrm{H}(E_R)|^2=1$, 
while for heavier elements we consider the simple exponential form factor \cite{Gould:1987ir,
eder, Wikstrom:2009kw}:
\be
 |F_i(E_R)|^2=\exp(-E_R/E_i), \quad \textrm{with}\quad E_i=3/(2 m_{N_i} R_i^2)\,,\quad
R_i= [ 0.91 \left( {m_{N_i}}/{\rm{GeV}} \right)^{1/3} + 0.3] \,\, \rm{fm} \,,
\ee
which has the advantage of making possible a simple analytical integration
of Eq.~(\ref{omegaminus}), to get 
\be
\Omega_{v_e, i}^{-}(w)=\frac{\sigma_i n_{N_i}}{w}\frac{(\mu_i+1)^2}{2m_\chi \mu_i} E_i \left[ e^{-m_\chi u^2/(2E_i)}
-e^{-m_\chi w^2 \mu_i/(2 \mu_{+,i}^2E_i)} \right]\,.
\label{Omegai}
\ee
We checked that using the more accurate Helm-Lewin-Smith form factor \cite{Lewin:1995rx,Duda:2006uk}, the capture rate would differ by less than 2\% in the mass
range considered, and the corresponding number of DM particles (to be discussed in the next section) by less 
than 1\%, for $m_\chi\leq 10$ GeV.

\begin{figure}[t]
\centering
\includegraphics[scale=1]{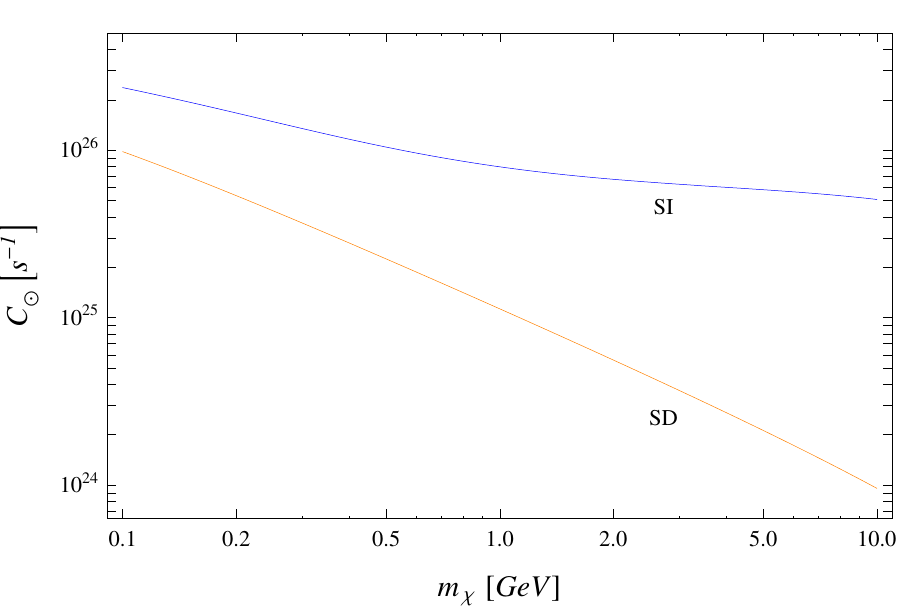}
\caption{\emph{{\small
The capture rate $C_\odot$  for $\sigma_p=10^{-40} \rm{cm}^2$ 
for SD  and SI.
}}}
\label{fig:Capture_rate}
\end{figure}

Finally, the total capture rate inside the Sun is obtained by integrating Eq.~(\ref{dCdV}), with (\ref{fvsun}), (\ref{OmegaH}) and (\ref{Omegai}), over the solar volume and summing over the different nuclear species 
in the Sun
\be
C_{\odot}=\sum_i \int_{\textrm{Sun}}  \frac{d C_{\odot,i} }{dV}\de^3 r,
\ee
where $i$ refers to the nucleus $i$.  
The quantity of phenomenological interest is
the DM-proton scattering cross section $\sigma_p$, which is related to the 
 cross section $\sigma_i$ on the nucleus $i$ (with mass number $A_i$) by
\be
\sigma_i=\sigma_p A_i^2\frac{m_{N_i}^2}{m_p^2}\frac{(m_\chi+m_p)^2}{(m_\chi+m_{N_i})^2}\,,
\ee
and we assume equal couplings 
of the DM to protons and neutrons. The generalization to account for different DM-nucleon couplings is straightforward.
For spin-independent (SI) DM-nucleus interactions, we have an enhancement of the cross section from constructive interference between nucleons inside the nucleus $i$. Therefore, we have included contributions from the most important elements up to Ni. On the other hand, for spin-dependent (SD) interactions, only Hydrogen  is considered since another dominant element, Helium, has  spin zero. 
So,   the capture rate for SD interactions is computed using Eqs.~(\ref{dCdV}) and (\ref{OmegaH}) with $\sigma_{\rm H}=\sigma_p$ and unit form factor $|F_{\rm H}(E_R)|^2=1$.
The capture rates we obtained are shown in Fig.~\ref{fig:Capture_rate}, for the SD and SI cases.
Our results are in very good agreement with those of Ref.~\cite{kw}.

We find the following simple fitting functions
for the capture rate corresponding to SD and SI DM-nucleus interactions
\bea
C_\odot&\simeq&3.57 \, e^{1.34\left[ \log\left(\frac{20 \GeV}{m_{\chi}}\right) \right]^{0.86}}
\left(\frac{ \sigma_p } {10^{-40} \rm{cm}^2}\right)
\times 10^{23}\,\textrm{s}^{-1}\,,
\qquad\qquad \textrm{(SD)}
\label{CratefitSD}\\
C_\odot &\simeq& 5.27 \, e^{3.73\times 10^{-2}\left[ \log\left(\frac{20 \GeV}{m_{\chi}}\right) \right]^{2.23}}
\left(\frac{ \sigma_p } {10^{-40} \rm{cm}^2}\right)
\times 10^{25}\,\textrm{s}^{-1}\,,\qquad \textrm{(SI)}
\label{CratefitSI}
\eea
valid in the range $0.1\leq m_\chi\leq 10$ GeV, with an accuracy better than 3\% and 6\%, respectively. 

\subsection{Evaporation}

As highlighted in the previous subsection, the formalism for describing evaporation
is identical to that for capture.
However, contrarily to capture, the evaporation is highly sensitive to the temperature of the distribution of nuclei
in the Sun, and therefore we now need to work in the finite temperature regime $T_{N_i}\neq 0$.
Also, we willl work in the regime where the Sun is optically thin with respect to the DM particles, and we do not consider the refinements of the calculations in the optically thick regime
\cite{gould90, gilliland}

The evaporation rate per unit volume at radius $r$ is given by
\be
\frac{\de E_{\odot, i}}{\de V}= \int^{v_e}_0 f_\odot(w)\Omega^{+}_{v_e, i}(w)\de w,
\label{dEdV}
\ee
with $\Omega^+_{v_e, i}$ given by Eq.~(\ref{omegaplus}).
We will approximate the
velocity distribution  $f_\odot(w)$ of the population of DM particles trapped in
the Sun, as  a Maxwell-Boltzmann distribution, 
depending on the DM mass and temperature
\be
f_\odot(w)=n_\chi{4\over \sqrt{\pi}}\left(\frac{m_\chi}{2T_\chi}\right)^{3/2} {w^2} e^{-m_\chi w^2/(2 T_\chi)}\,.
\label{MB2}
\ee
The approximation of thermal distribution is valid for DM mass $m_\chi\sim 1$ GeV, while the actual DM distribution deviates
from the thermal distribution for larger masses.
We use the results of \cite{gould87} 
to account for the corrections due to a non-thermal distribution.
The total evaporation rate per DM particle is obtained by integrating Eq.~(\ref{dEdV}) over the
solar volume and divide by the total number of DM particles in the Sun
\be
E_\odot=\frac{\sum_i \int_{\textrm{Sun}}  \frac{d E_{\odot, i} }{dV}\de^3 r}
{\int_{\textrm{Sun}} n_\chi(r) \, \de^3 r}
\label{eq:evapo_per_DM}
\ee
 Again, for SD interactions, only Hydrogen is considered but for SI interactions, we include all
 the  elements up to Nickel, using the solar model AGSS09 \cite{solar}.
 
\begin{figure}[t]
\centering
\includegraphics[scale=0.8]{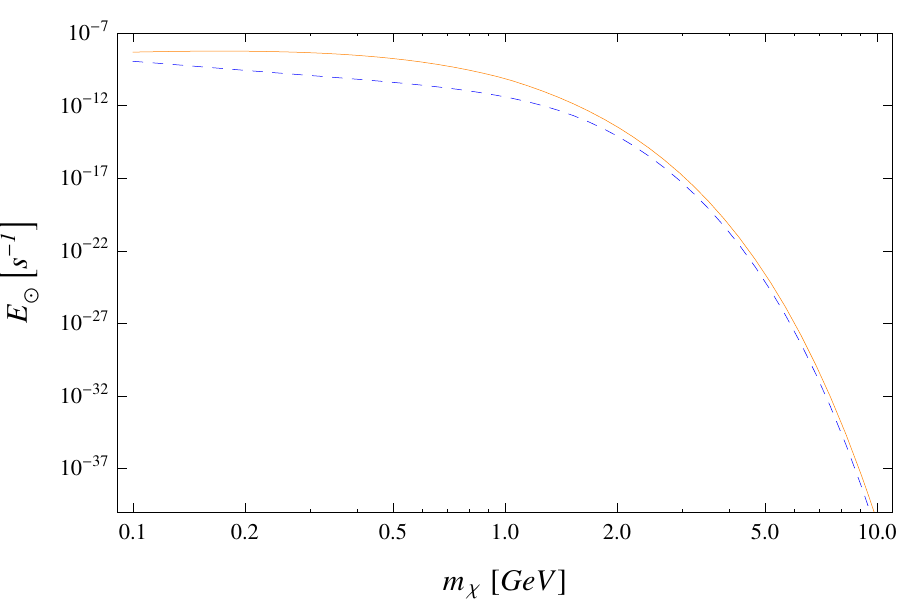}
\hspace{0.3cm}
\includegraphics[scale=0.8]{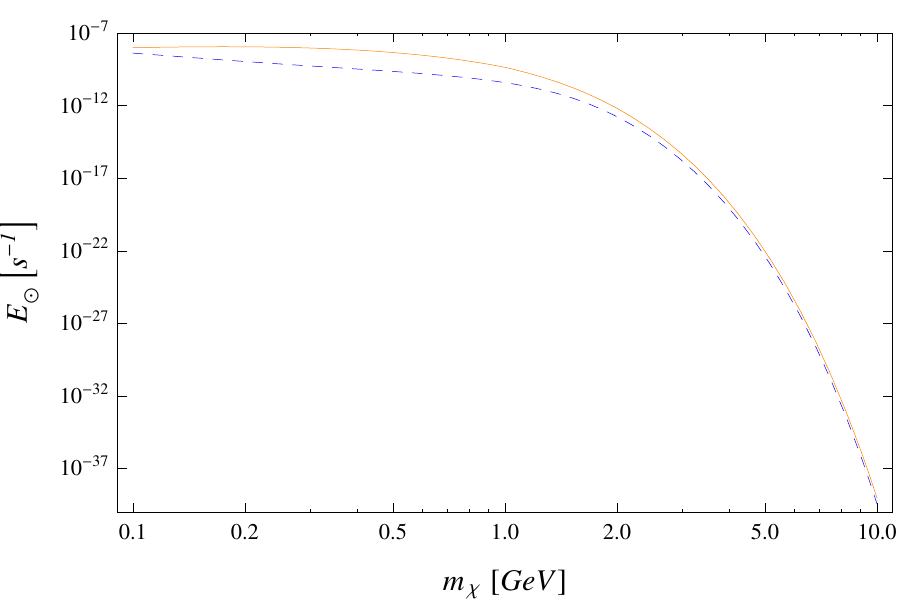}
\caption{\emph{{\small
The evaporation rate $E_\odot$, for $\sigma_p=10^{-40} \rm{cm}^2$
(orange solid line) for SD \emph{(left panel)} and SI \emph{(right panel)}. 
We compare it to the simple
analytical approximation $E_\odot^{\rm approx}$  in  Eq.~(\ref{eq:evapo_per_DM_approx})
(blue dashed line).
}}
}
\label{fig:Evaporation_comparison}
\end{figure}

There is a simple analytical approximation of Eq.~(\ref{eq:evapo_per_DM}) 
 \cite{gould87, gould90}, which is valid for $m_\chi/m_N>1$
\be
E_\odot^{\rm approx}\simeq {8\over \pi^3}\sqrt{\frac{2 m_\chi}{\pi T_\odot(\bar r)}}
\frac{v_e(0)^2}{\bar r^3}
e^{-\frac{m_\chi v_e(0)^2}{2 T_\odot(\bar r)}}\Sigma_{\rm evap},
\label{eq:evapo_per_DM_approx}
\ee
where $v_e(0)$ is the escape velocity at the solar center.
The quantity $\Sigma_{\rm evap}$ is  the sum of the
scattering cross sections of all the nuclei within a radius $r_{95\%}$, where the
solar temperature has dropped to 95\% of the DM temperature. The derivation of Eq.~(\ref{eq:evapo_per_DM_approx}) is sketched in  Appendix
\ref{app:Eapprox}. We present our numerical results for $E_\odot$ in Fig.~\ref{fig:Evaporation_comparison}. 
Notice that for $m_\chi\gtrsim 4$ GeV the evaporation rate drops rapidly. 
We found that the approximated
formula $E_\odot^{\rm approx}$ of Eq.~(\ref{eq:evapo_per_DM_approx}) is off by a factor $\lesssim 4$ 
with respect to the full numerical result,
in the relevant region
$2\lesssim m_\chi\lesssim 5$ GeV, in agreement with what stated in Ref.~\cite{gould90}.

For the evaporation rate  for SI and SD DM-nucleus interactions,
 we find the following simple fitting functions
\bea
E_\odot&\simeq&1.09 \,  e^{-34.97\left(\frac{1 \GeV}{m_\chi}\right)^{0.0467}
-9.25 \left(\frac{m_\chi}{1 \GeV}\right)^{0.95}}\left(
\frac{ \sigma_p } {10^{-40} \rm{cm}^2}\right) \times 10^{9}\,\rm{s}^{-1}\,,
\qquad\quad \textrm{(SD)}
\label{EratefitSD}\\
E_\odot&\simeq&5.13 \,  e^{-39.6\left(\frac{1 \GeV}{m_\chi}\right)^{0.077}
-8.92 \left(\frac{m_\chi}{1 \GeV}\right)^{0.97}}\left(
\frac{ \sigma_p } {10^{-40} \rm{cm}^2}\right) \times 10^{11}\,\rm{s}^{-1}\,,
\qquad\quad\; \textrm{(SI)}
\label{EratefitSI}
\eea
which reproduce the full numerical results with an accuracy better than $14\%$ and $10\%$, respectively,
in the range $0.5\leq m_\chi\leq 8$ GeV. For heavier DM masses, the evaporation is completely
negligible.

\section{Results}
\label{sec:number}

\subsection{The number of DM particles in the Sun}

We have now all the tools to determine the number  of DM particles in the Sun,
which depends on the DM mass $m_\chi$, its annihilation cross section $\langle\sigma_{\rm ann}v\rangle_\odot$ 
and its scattering cross section with proton $\sigma_p$.
 The time evolution of the number  $N$ is described by the simple differential equation
 \cite{Jungman:1995df}
\be
\frac{\de{N(t)}}{\de t}=C_\odot  - E_\odot N(t) - A_\odot N(t)^2\,,
\label{dNdt}
\ee
whose solution, evaluated at the age of the Sun $t_\odot$,  is
\be
 N(t_\odot)=\sqrt{C_\odot\over A_\odot}\cdot\frac{\tanh(k t_{\odot}/\tau)}{k + \frac{1}{2}E_\odot\tau
\tanh(k t_{\odot}/\tau)}\equiv N_\chi,
\ee
where $\tau\equiv1/\sqrt{C_\odot A_\odot}$ and $k\equiv \sqrt{1+(E_\odot \tau/2)^2}$.

Depending on the DM mass and cross sections, the different processes have different relevances, 
and ultimately two regimes are possible: capture and annihilation are in equilibrium, or capture and evaporation
are in equilibrium.
For the cross sections of interest, $\sigma_p\gtrsim 10^{-42} \cm^2$,
 the quantity $k t_\odot/\tau$ is always bigger than one meaning that the equilibrium condition is always fulfilled.
When evaporation is negligible, $E_\odot \tau\ll 1$, then $k\simeq 1$ and the number $N_\chi$ simply reduces to
\be
N_\chi\simeq \sqrt{C_\odot\over A_\odot}\, \tanh( t_{\odot}/\tau)\simeq \sqrt{C_\odot\over A_\odot}\,.
\ee
In this situation the capture and annihilation processes are in equilibrium.
On the other hand, in the opposite regime  $E_\odot \tau\gg 1$, the annihilation
becomes negligible and the equilibrium is attained by capture and evaporation
and the number of DM particles becomes  
\be
N_\chi\simeq\frac{C_\odot}{E_\odot}
\label{NCE}
\ee
becomes independent of the DM-nucleus cross section.

The main parameter determining whether evaporation is relevant or not is the DM mass.
Since the evaporation drops rapidly for about $m_\chi\gtrsim 4$ GeV, so we expect that in this regime
 annihilation and capture are in equilibrium;
however, for lighter DM, the  capture goes in equilibrium with evaporation.

\begin{figure}[t]
\centering
\includegraphics[scale=0.8]{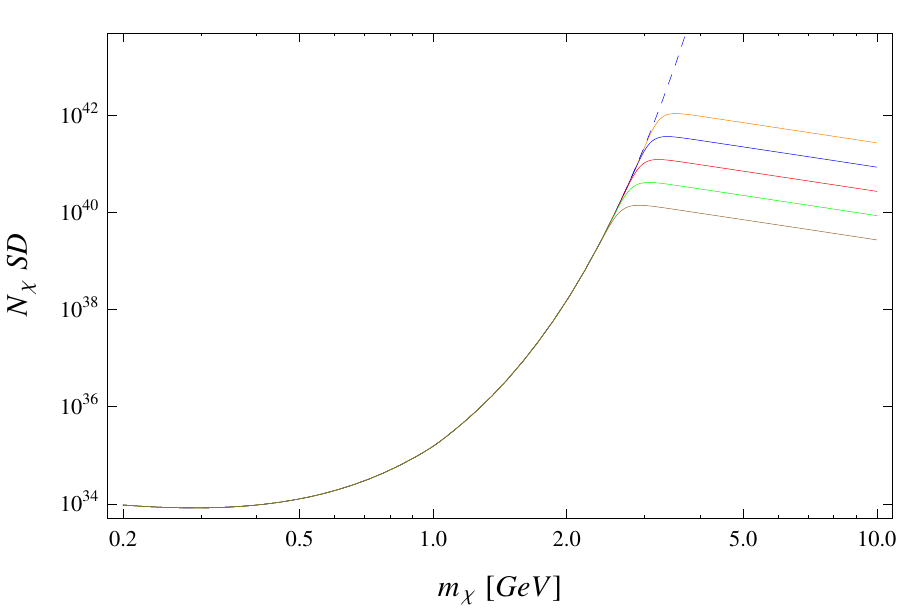}
\hspace{0.3cm}
\includegraphics[scale=0.8]{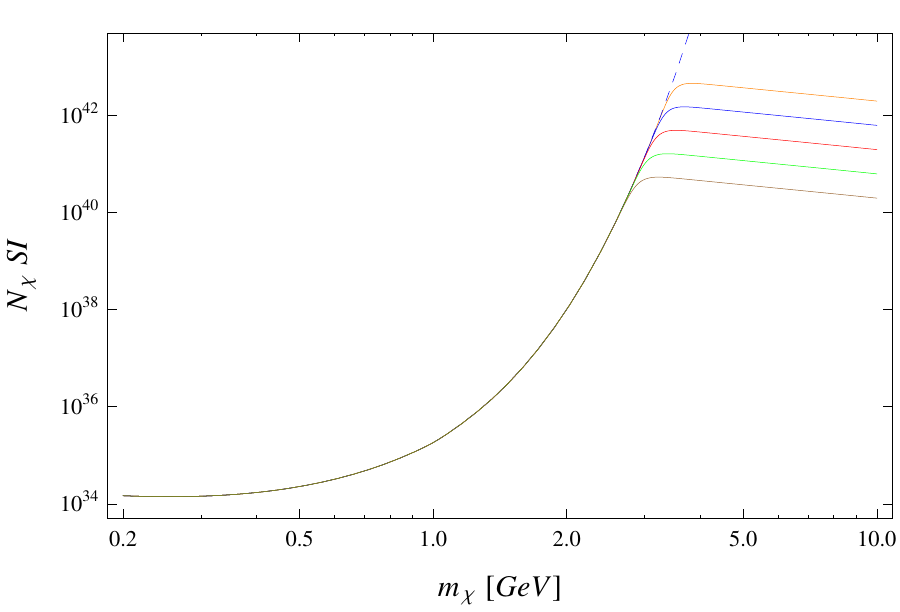}
\caption{\emph{{\small
The number of DM particles  in the Sun $N_\chi$, as a function of the DM mass,
 for SD \emph{(left panel)} and SI \emph{(right panel)},
 for different values of the DM-proton cross section $\sigma_p=10^{-40}, 10^{-39},
 10^{-38}, 10^{-37}, 10^{-36} \cm^2$ (from
 bottom to top).
 We have set $\langle\sigma_{\rm ann}v\rangle_\odot=3\times 10^{-26}\rm{cm}^3/s$.
 We also show  the curve (blue dashed line) corresponding to $C_\odot/E_\odot$, which $N_\chi$
 tends to when the evaporation-capture equilibrium is reached.
}}}
\label{fig:NSISDCE}
\end{figure}

In Fig.~\ref{fig:NSISDCE}, we show $N_\chi$ as a function of $m_\chi$ for different values of $\sigma_p$, in the
range $10^{-40}\div 10^{-36} \rm{cm}^2$. Notice that $N_\chi$ tends to the curve $C_\odot/E_\odot$,
corresponding to when the equilibrium between capture and evaporation is attained, and the
number of DM particles does not depend on $\sigma_p$ anymore.
Notice also that the maximum of $N_\chi$ occurs around $m_\chi \sim 3$ GeV because below this value the evaporation is important,  yielding fewer $N_\chi$, and above that  the number of DM particles passing through the Sun decreases as $\rho_\odot/m_\chi\simeq 0.3 \GeV  \cm^{-3}/m_\chi$.

\subsection{The minimum testable DM mass}

In order to characterize how evaporation affects $N_\chi$, one can define two quantities with dimension
of a mass:
the ``evaporation'' mass $m_{\rm{evap}}$  and the ``minimum'' mass $m_{\rm min}$.
First, the evaporation mass is defined as the mass for which the inverse of the evaporation rate is equal to the age of the Sun $t_\odot\simeq 4.7 \times 10^9$ yrs, i.e., $E_\odot(m_{\rm{evap}})\equiv 1/t_\odot$
\cite{gould90}. 
Second, we introduce
the mass $m_{\rm min}$, corresponding to the DM mass for which $N_\chi$ approaches the equilibrium value
$C_\odot/E_\odot$, which is independent of $\sigma_p$. Quantitatively, 
\be
\left\vert N_\chi(m_{\rm min})-\frac{C_\odot}{E_\odot} \right\vert \equiv 0.1 \, N_\chi(m_{\rm min})\,,
\ee
where we arbitrarily chose 10\% as a satisfactory level of $N_\chi$ approaching  $C_\odot/E_\odot$.
The standard lore regarding $m_{\rm{evap}}$ is that 
the evaporation rate becomes negligible when $m_\chi \geq m_{\rm{evap}}$; on the other hand, for $m_\chi \leq m_{\rm{evap}}$,  the annihilation rate becomes negligible. As a consequence,  one would expect that for $m_\chi\leq m_{\rm evap}$ the
capture and evaporation processes are in equilibrium and the number of DM
particles in the Sun can be approximated by the equilibrium value in Eq.~(\ref{NCE}), which does not depend on $\sigma_p$.
What we want to point out here is that  it is actually $m_{\rm min}$ (and not $m_{\rm evap}$) which qualifies the 
inability of extracting constraints on $\sigma_p$, 
since  the number of DM particles $N_\chi$ is not sensitive to $\sigma_p$ anymore,
for  $m_{\chi}\leq m_{\rm min}$.

We have found some simple fits of of $m_{\rm min}$  as a function of $\sigma_p$ and $\langle\sigma_{\rm ann}v\rangle_\odot$
\bea
m_{\rm min}&\simeq&\left[2.5  +0.15 \log_{10} \left( \frac{\sigma_p}{10^{-40}\,\rm{cm^2}}  \right) 
-0.15 \log_{10} \left( \frac{\langle\sigma_{\rm ann}v\rangle_\odot}{3\cdot 10^{-26}\,\rm{cm^3}/s}  \right) 
\right]\GeV
\;(\rm{SD})\,,
\label{mminSD}\\
m_{\rm min}&\simeq&\left[2.7  +0.15 \log_{10} \left( \frac{\sigma_p}{10^{-40}\,\rm{cm^2}}  \right) 
-0.15 \log_{10} \left( \frac{\langle\sigma_{\rm ann}v\rangle_\odot}{3\cdot 10^{-26}\,\rm{cm^3}/s}  \right) 
\right]\GeV
\;(\rm{SI}),
\label{mminSI}
\eea
which are valid to better than $1\%$, in the interval: $10^{-42} \cm^2\leq \sigma_p \leq 10^{-30}\cm^2$, 
 $3\times 10^{-27} \cm^3/\textrm{s}\leq\langle\sigma_{\rm ann}v\rangle_\odot\leq 3\times 10^{-25} \cm^3/\textrm{s}$.
In these intervals, the evaporation mass is always greater than $m_{\rm min}$.

A simple argument to understand the positive correlation between $m_{\rm min}$ and $\sigma_p$ 
goes as follows.
First of all, in the regime where capture and evaporation are the relevant processes, the larger $m_\chi$,
the more difficult is for nuclei to expel DM particles, so $N_\chi$ is larger. 
Then, increasing $\sigma_p$ leads to more DM particles captured by the Sun, so larger $N_\chi$.
Therefore, $m_{\rm min}$ turns out to be larger.

\begin{figure}[t!]
\centering
\includegraphics[scale=0.6]{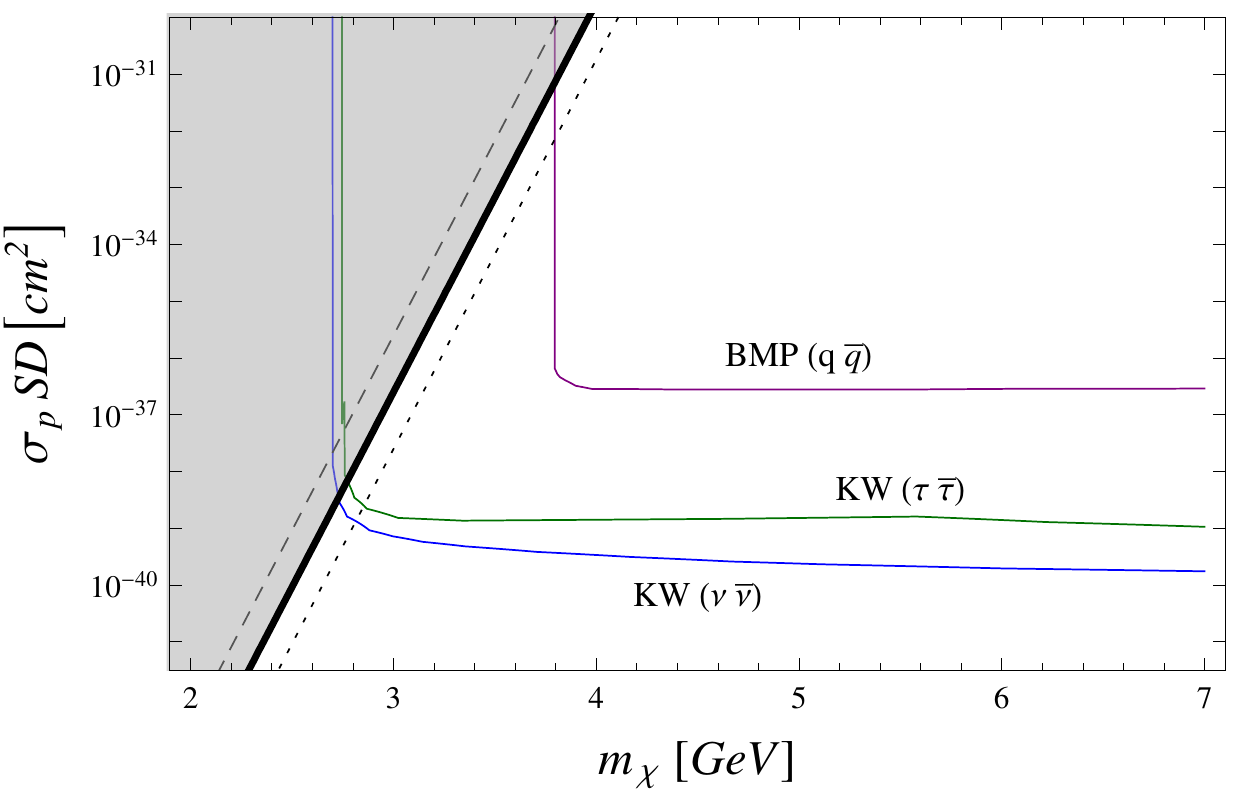}
\hspace{0.7cm}
\includegraphics[scale=0.6]{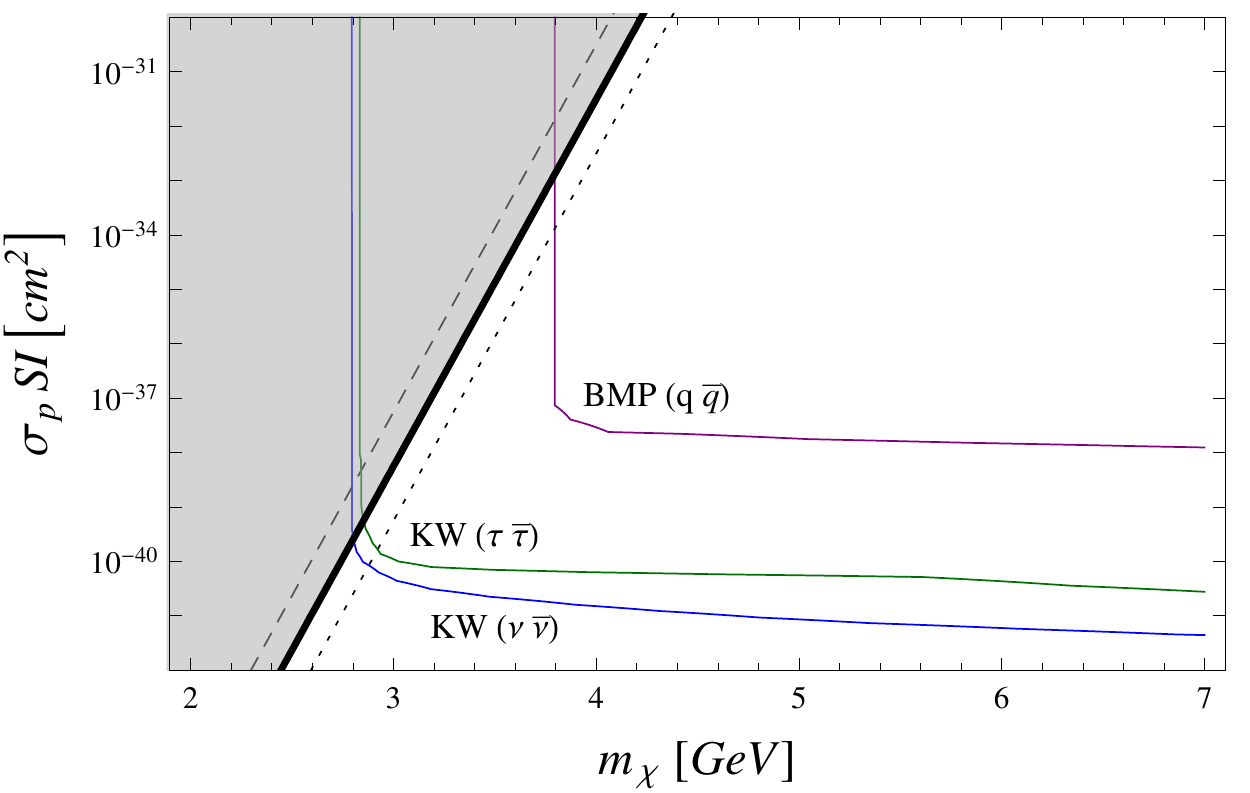}
\caption{\emph{{\small
The region in the $(m_\chi,\sigma_p)$ plane \emph{(shaded area)} which is not testable by detectors of neutrinos from the Sun, as discussed in the text, for SD \emph{(left panel)} and SI \emph{(right panel)} interactions.
The thick black line corresponds to $m_{\rm min}$ for the reference annihilation cross section $\langle\sigma_{\rm ann} v\rangle_\odot=
3\times 10^{-26}$ cm$^3$/s,  the dotted (dashed) black lines correspond to $\langle\sigma_{\rm ann} v\rangle_\odot=0.1\, (10)$ times the reference value.
For comparison, we also show the exclusion curves obtained in Ref.~\cite{Bernal:2012qh} (BMP) for ${\rm DM \, DM}\to q\bar q$, and in Ref.~\cite{kw} (KW) for  ${\rm DM \, DM}\to \tau\bar \tau, \nu\bar\nu$.
}}
}
\label{fig:BMP_mce}
\end{figure}

In Fig.~\ref{fig:BMP_mce}, we  plot  $m_{\rm min}$ in the $(\sigma_p, m_\chi)$ plane. The plot shows the region  of parameter space where it is not possible to contrain $\sigma_p$ with neutrino
data from the Sun. For comparison,
we also show some of the exclusion curves obtained in the analysis of  Super-K data 
of Refs.~\cite{Bernal:2012qh, kw}.
For instance, for $m_\chi\lesssim 4 \GeV$,  data on neutrinos from the Sun  are not able to provide information on the DM-proton scattering cross section  below $\sigma_p\lesssim 10^{-31}\cm^2$.
On the other hand, for a given value of the scattering cross section  
there is a minimum DM mass (see Eqs.~(\ref{mminSD})-(\ref{mminSI})) which can be probed
by neutrino fluxes from the Sun.
Increasing (decreasing) the annihilation cross section  $\langle\sigma_{\rm ann} v\rangle_\odot$ leads to a smaller (bigger) $m_{\rm min}$ at fixed $\sigma_p$, as confirmed
by Eqs.~(\ref{mminSD})-(\ref{mminSI}); the effect of varying the annihilation cross section
by a factor of 10 with respect to its reference value is also shown in Fig.~\ref{fig:BMP_mce}.

\section{Conclusions}
\label{sec:conclusions}

In this paper we have considered the implications of the presence of GeV-scale DM
in the Sun, the relevant processes it is subject to, and the 
 constraints which can be placed on its properties, namely mass and cross sections, 
 using neutrino data.
We can summarize our main results as follows:
\begin{itemize}
\item 
for DM masses below about 4 GeV the effect of evaporation cannot be neglected, and we provide handy and accurate fitting functions for all the relevant processes of light DM in the Sun: annihilation Eq.~(\ref{Aratefit}), capture Eqs.~(\ref{CratefitSD})-(\ref{CratefitSI}) and evaporation
Eqs.~(\ref{EratefitSD})-(\ref{EratefitSI});
\item we point out a limitation on extracting cross section bounds when evaporation is important;
we provide expressions for the minimum DM mass below which the number of
DM particles in the Sun
does not depend on $\sigma_p$, Eqs.~(\ref{mminSD})-(\ref{mminSI}), and
the link with DM direct detection bounds disappears;
\item we identify the region of the parameter space ($m_\chi, \sigma_p, \langle\sigma_{\rm ann} v\rangle_\odot$) (see Fig.~\ref{fig:BMP_mce}) which is not accessible by data on  neutrino 
fluxes from the Sun.
\end{itemize}

\section*{Acknowledgments}
We thank P.~Machado, S.~Palomares-Ruiz and S.~Petcov for useful discussions and comments on the manuscript.
We also thank J.~Edsj\"o for interesting discussions.
ADS acknowledges partial support from the  European Union FP7  ITN INVISIBLES (Marie Curie Actions, PITN-GA-2011-289442).
GB is partially supported by the Swiss National
Science Foundation (SNSF), project N. 200021140236.

\appendix

\section{Analytical calculation of scattering rate}
\label{app:A}

In this appendix we present the calculation of the scattering rate per unit time $R^\pm(w\to v)$
at which a single DM particle of velocity $w$ scatters to a final velocity between $v$ and $v+\de v$,
off a thermal distribution of nuclei  with number density  $n_{N}$, mass $m_N$ and temperature $T_N$ (for simplicity, throughout this appendix we drop the index $i$ referring to a particular nucleus).
We will use this result to  compute the rates for capture and evaporation. This calculation was first performed by Gould in Ref.~\cite{gould87} and we reproduce it here, although in a different
form.
The differential scattering rate of a DM particle of initial speed $w$ (in the lab frame) and final speed $v$ on a nucleus of speed $u$ is
\be
C t^2s e^{-\kappa^2(2\mu \mu_{+}t^2+2\mu_{+}s^2)}\theta (w-\vert s-t\vert) \theta (s+t-w) \delta{[v-(s^2+t^2-2zst)^{1/2}]}\mathrm{d}z\mathrm{d}s\mathrm{d}t\,,
\ee
where $s$, $t$ and $\kappa$ are defined as
\bea
(1+\mu)s=\vert\vec{u}+\mu \vec{w}\vert\,,\qquad
(1+\mu)t=\vert\vec{w}-\vec{u}\vert\,,\qquad
\kappa=\sqrt{\dfrac{m_{N}}{2 T_{N}} }\,,
\eea
$\mu, \mu_{\pm}$ are defined in (\ref{alphabetamu}) and $C$ is a multiplicative factor
\be
C = \frac{16\mu_{+}^4}{\sqrt{\pi}}\kappa^3 n_N \sigma \frac{e^{\kappa^2\mu w^2}}{w}\,.
\ee
The momentum conservation in the lab frame is obtained by integrating the following expression
\be
\int_{-1}^1 \! \delta [v-(s^2+t^2-2zst)^{1/2}] \, \mathrm{d}z=\frac{v}{st}\theta (v-\vert s-t\vert) \theta (s+t-v)\,.
\ee
The integration domain is determined by the 4 $\theta$-functions, that give us 4 inequalities
\bea
v-\vert s-t\vert \ge 0,	\quad  w-\vert s-t\vert \ge 0\,,\\
s+t-v\ge 0,\quad  s+t-w \ge 0\,,
\eea
and we get that the new variables are subject to the constrains
\be
x_1=\frac{|v-w|}{2}\, ,\qquad  x_2=\frac{v+w}{2} \,,
\ee
\be
\begin{cases} x_1 \le t \le x_2\,, \quad \max[v,w]-t \le s \le \min[v,w]+t \,,\\ 
x_2 \le t \le \infty\,, 
\quad t-\min[v,w] \le s \le \min[v,w]+t\,. \end{cases}
\ee
The $s$ integral is gaussian, thus we obtain (case $v>w$)
\bea
R^+(w\to v)&=&C v \int_{x_1}^{x_2}  \mathrm{d}t \int_{v-t}^{w+t} \! \mathrm{d}s\,
 t e^{-\kappa^2(2\mu \mu_{+}t^2+2\mu_{+}s^2)} \, 
+C v \int_{x_2}^\infty  \mathrm{d}t \int_{t-w}^{w+t} \! \mathrm{d}s\, t e^{-\kappa^2(2\mu \mu_{+}t^2+2\mu_{+}s^2)} \,\nn\\
&=&\frac{C v}{\kappa\sqrt{2\mu_{+}}}\int_{x_1}^{x_2} \! \mathrm{d}t \chi({\kappa\sqrt{2\mu_{+}}(v-t)},{\kappa\sqrt{2\mu_{+}}(w+t)}) t e^{-\kappa^2(2\mu \mu_{+}t^2)} \, \nn\\
&&+\frac{C v}{\kappa\sqrt{2\mu_{+}}}\int_{x_2}^\infty \! \mathrm{d}t \chi({\kappa\sqrt{2\mu_{+}}(t-w)},{\kappa\sqrt{2\mu_{+}}(w+t)}) t e^{-\kappa^2(2\mu \mu_{+}t^2)} \, ,
\eea
where
\be
\chi(a,b)\equiv\int_a^b \de y e^{-y^2}=\frac{\sqrt{\pi}}{2}\left[
\textrm{Erf}(b)-\textrm{Erf}(a)
\right]\,.
\label{chi}
\ee
Using the fact that, for any real numbers $a, b, c, d, e, A$,
\bea
\int  \mathrm{d}t \chi({b t + c},{d t + e}) t e^{-A^2 t^2} &=& -\frac{e^{-A^2 t^2}\chi({b t + c},{d t + e})}{2A^2} 
-\frac{\sqrt{\pi}b}{4 A^2 }\frac{e^{-\frac{A^2 c^2}{A^2+b^2}}\textrm{Erf}[\frac{b c + A^2 t + b^2 t}{\sqrt{A^2+b^2}}]}{\sqrt{A^2+b^2}}\nn\\
&&+\frac{\sqrt{\pi}d}{4 A^2 }\frac{ e^{-\frac{A^2 e^2}{A^2+d^2}}\textrm{Erf}[\frac{d e + A^2 t + d^2 t}{\sqrt{A^2+d^2}}]}{\sqrt{A^2+d^2}}\,,
\eea
and defining $\alpha_{\pm}, \beta_{\pm}$ as in (\ref{alphabetamu}), we get the final result
\bea
R^{+}(w\to v) =
\frac{2}{\sqrt{\pi}} n_N \sigma \frac{v}{w} \frac{\mu_{+}^2}{\mu}\left[\chi({\alpha_{-}},{\alpha_{+}})+e^{-k^2\mu (v^2-w^2)}\chi({\beta_{-}},{\beta{+}})\right] \,.
\label{Rplus}
\eea
The case $w>v$ can be done in the same way, and we get
\bea
R^{-}(w\to v) = 
\frac{2}{\sqrt{\pi}} n_N \sigma \frac{v}{w} \frac{\mu_{+}^2}{\mu}\left[\chi({-\alpha_{-}},{\alpha_{+}})+e^{-k^2\mu (v^2-w^2)}\chi({-\beta_{-}},{\beta_{+}})\right]\,.
\label{Rminus}
\eea
The results in  Eqs.~(\ref{Rplus})-(\ref{Rminus}) reproduce the expression in Eq.~(\ref{Rplusminus}).

\section{Analytical approximation of the evaporation rate}
\label{app:Eapprox}

Using the identities in  Ref.~\cite{gould87} to evaluate the integrals (\ref{omegaminus})-(\ref{omegaplus}),
one finds (for simplicity, throughout this appendix we drop the index $i$ referring to a particular nucleus)
\bea
\Omega^{\pm}_{v_e}(w)&=&\pm \frac{1}{ 2\sqrt{\pi} } \frac{ 2T_{N} }{m_{N}} \frac{1}{\mu^2} \frac{\sigma n_{N}}{w}
\left[ \mu\left(\pm\alpha_{+} e^{- \alpha^2_{-} }-\alpha_{-}e^{- \alpha^2_{+} }  \right)\right. \nn\\
&& \left.
+ \left( \mu-2\mu \alpha_{+}\alpha_{-} - 2 \mu_{+}\mu_{-}  \right) \chi(\pm\alpha_{-},\alpha_{+}) 
  + 2 \mu^2_{+}  e^{ -\frac{m_\chi  ( v_e^2 -w^2)}{2T_{N}}}\chi(\pm\beta_{-},\beta_{+})   \right]\,,
\label{Omegapm}
\eea
where $\chi(a,b)$ is defined as in (\ref{chi}), 
and the evaporation rate per unit volume is defined as in  Eq.~(\ref{dEdV})
\be
\frac{\de E_{\odot}}{\de V}= \int^{v_e}_0 f_0(w)\,\Omega^{+}_{v_e}(w)\de w\,.
\ee
This is a function of $r, m_\chi, \sigma$.
The analytical evaluation of this integral is possible (although lengthy)
when  $f_0$ is a thermal Maxwell-Boltzmann distribution as in (\ref{MB2}), and the result is
\be
\frac{\de E_{\odot}}{\de V}=\sigma A(r,m_\chi)\,
n_N(r)n_0 e^{-m_\chi\phi(r)/T_\chi} e^{-( E_{\rm esc}(r)- E_{\rm esc}(0))/T_\chi}
{E_{\rm esc}(r)\over E_{\rm esc}(0)} \widetilde{R}(m_\chi)\,,
\label{RR}
\ee
where $E_{\rm esc}(r)=(1/2)m_\chi v_e(r)^2$ is the escape energy at radius $r$, and
\bea
\widetilde{R}(m_\chi)&=& {2\over \sqrt{\pi}} \sqrt{2T_\chi\over m_\chi}
{E_{\rm esc}(0)\over T_\chi}e^{-E_{\rm esc}(0)/T_\chi}\,,
\label{Rtilde}\\
A(r,m_\chi)&=&\frac{1}{\sqrt{\pi}}\left(\frac{T_N}{T_\chi}\right)^{3/2}
\left\{
e^{-\frac{E_{\rm esc}(r)}{T_\chi}\frac{\mu T_N/T_\chi}{\mu_-^2+\mu (T_N/T_\chi)}}
\left[\frac{T_\chi}{T_N}\frac{\mu_-}{\sqrt{\mu_-^2+\mu T_N/T_\chi}}
\left(1+\frac{\mu_-^2}{\mu T_N/T_\chi}-\frac{\mu_-^2}{\mu}\right)
\right.\right.\nn\\
&&
\qquad\left.
+\frac{\mu_+^3}{\mu\sqrt{\mu_-^2+\mu T_N/T_\chi}\left(\frac{T_N}{T_\chi}-1\right)}
\right]\chi(\gamma_-, \gamma_+)\nn\\
&&
+\frac{T_\chi}{T_N}\left[
\left(\frac{E_{\rm esc}(r)}{T_N}-\frac{1}{2\mu}+\frac{\mu_-^2}{\mu}\left(1-\frac{T_\chi}{T_N}\right)\right)
\chi(\alpha_-,\alpha_+)
-\frac{\mu_+^2}{\mu}\frac{1}{1-\frac{T_\chi}{T_N}}\chi(\beta_-,\beta_+)\right.\nn\\
&&\left.\left.
\qquad
+2v_e(r)\sqrt{\frac{m_N}{2T_N}}\left(e^{-\mu^2 m_N v_e^2 /(2T_N)}-
\mu e^{-m_N v_e^2 /(2T_N)}\right)
\right]
\right\}\,,
\label{Afunction}\\
\gamma_\pm&\equiv&\sqrt{\frac{m_N}{2T_N}}v_e
[\sqrt{\mu_-^2+\mu T_N/T_\chi}
\pm{\mu_-^2}/{\sqrt{\mu_-^2+\mu T_N/T_\chi}}]\,.
\eea
For convenience, we kept separated in (\ref{RR}) the $r$-dependent and $r$-independent terms.
\bea
E_\odot(m_\chi, \sigma)&=&{\sigma\widetilde{R}(m_\chi) \over 
\int_{\textrm{Sun}} e^{-m_\chi\phi(r)/T_\chi} \de^3 r}
\int_{\textrm{Sun}}  \de^3 r
\left[
A(r,m_\chi)\,
n_N(r) e^{-m_\chi\phi(r)/T_\chi} e^{-( E_{\rm esc}(r)- E_{\rm esc}(0))/T_\chi}
{E_{\rm esc}(r)\over E_{\rm esc}(0)}\right]\nn\\
&\equiv&{\sigma\widetilde{R}(m_\chi) \over \int_{\textrm{Sun}} e^{-m_\chi\phi(r)/T_\chi} \de^3 r}
\times I(m_\chi)\,.
\label{evap1}
\eea
This is the most general result for the evaporation rate, where the functions $\widetilde R$ and $A$
are given by Eqs.~(\ref{Rtilde}) and (\ref{Afunction}), respectively.
The underlying assumptions are:
isotropic and velocity-independent cross section, thermal distributions of DM and nuclei.

A simple  analytical approximation can be derived under the further hypothesis
that $\bar r$ is very small (corresponding to a rather large $m_\chi/m_N$).
In this regime, Ref.~\cite{gould87} argues  that the function $A$ can be approximated as 
\be
A(r,m_\chi)\simeq \theta(T_\odot(r)-0.95 T_\odot(\bar r))=\theta(T_\odot(r)-T_\odot(r_{95\%}))
=\theta(r_{95\%}-r)\,,
\label{Atheta}
\ee
being $T_\odot$ a monotonically decreasing function of $r$.
The radius $r_{95\%}$ is defined as the radius where
solar temperature has dropped to 95\% of the DM temperatrure, i.e.
\be
T_\odot(r_{95\%})=0.95 \, T_\odot(\bar r)\,,
\ee
and consequently, the number of nuclei within the radius $r_{95\%}$ is
\be
N_N^{95\%}=\int_{\rm Sun}\theta(r_{95\%}-r) n_N(r) \de^3 r\,.
\ee
The theta function (\ref{Atheta}) forces the integrand in $I(m_\chi)$ to be evaluated for very small region of $r$
close to the solar core, therefore the exponentials and the ratio of escape energies can be approximated
with 1. It only remains
\be
I(m_\chi)\simeq \int_{\rm Sun} \theta(r_{95\%}-r) n_N(r) \de^3 r = N_N^{\rm 95\%}\,.
\label{Iapprox}
\ee
This quantity, combined with the total DM-nucleus cross section  $\sigma$, gives
the evaporation cross section  $\Sigma_{\rm evap}$, which is  the sum of the
scattering cross sections of all the nuclei within a  radius $r_{95\%}$
\be
\Sigma_{\rm evap}=\sigma N_N^{95\%}\,.
\ee
Furthermore, within a small  region the density can be taken as constant $\rho_\odot=\rho_\odot(\bar r)$ and
the gravitational potential reads $\phi(r)=(2\pi/3)\rho_\odot r^2 G_N$. Thus, 
 the effective volume of DM  is simply
\be
\int_{\textrm{Sun}} e^{-m_\chi\phi(r)/T_\chi} \de^3 r={3\sqrt{3}\over 2\sqrt{2}}\left({T_\chi\over G_N \rho_\odot m_\chi }\right)^{3/2}\,,
\ee
while  the mean DM orbit radius, the mean DM velocity 
and the escape energy at the solar center  are
\be
\bar r= \sqrt{ 6 T_\odot(\bar r)\over \pi^2 G_N \rho_\odot(\bar r) m_\chi}\,,\qquad
\bar v= \sqrt{8T_\odot(\bar r)\over \pi m_\chi}\,,
\qquad
E_{\rm esc}(0)=\frac{1}{2} m_\chi v_e(0)^2\,.
\label{barv}
\ee
Finally, the evaporation rate (\ref{evap1}) can be re-written using Eqs.~(\ref{Iapprox})-(\ref{barv}),
\be
E_\odot\simeq 
E_\odot^{\rm approx}=
{8\over \pi^3}{ E_{\rm esc}(0) \bar v \over \bar r^3 T_\odot(\bar r)}
e^{-E_{\rm esc}(0)/T_\odot(\bar r)}\Sigma_{\rm evap}\,,
\ee
which recovers  Eq.~(\ref{eq:evapo_per_DM_approx}).



\begin{thebibliography}{99}

\bibitem{rott}	  
  C.~Rott, J.~Siegal-Gaskins and J.~F.~Beacom,
          \href{http://arXiv.org/abs/1208.0827}{[arXiv:1208.0827]}.      
  
\bibitem{Bernal:2012qh}
  N.~Bernal, J.~Martin-Albo and S.~Palomares-Ruiz,
        	  \href{http://arXiv.org/abs/1208.0834}{[arXiv:1208.0834]}.      

\bibitem{superk}
  T.~Tanaka {\it et al.}  [Super-Kamiokande Collaboration],
  Astrophys.\ J.\  {\bf 742}, 78 (2011)
      	  \href{http://arXiv.org/abs/1108.3384}{[arXiv:1108.3384]};
  S.~Desai {\it et al.}  [Super-Kamiokande Collaboration],
  Phys.\ Rev.\ D {\bf 70}, 083523 (2004)
  [Erratum-ibid.\ D {\bf 70}, 109901 (2004)]
      	  \href{http://arXiv.org/abs/hep-ex/0404025}{[hep-ex/0404025]};

\bibitem{Vogel:1999zy} 
  P.~Vogel and J.~F.~Beacom,
  Phys.\ Rev.\ D {\bf 60}, 053003 (1999)
      	  \href{http://arXiv.org/abs/hep-ph/9903554}{[hep-ph/9903554]};  
    
\bibitem{Yuksel:2007ac} 
  H.~Yuksel, S.~Horiuchi, J.~F.~Beacom and S.~'i.~Ando,
  Phys.\ Rev.\ D {\bf 76}, 123506 (2007)
          	  \href{http://arXiv.org/abs/0707.0196}{[arXiv:0707.0196]};    
    S.~Palomares-Ruiz and S.~Pascoli,
  Phys.\ Rev.\ D {\bf 77}, 025025 (2008)
            	  \href{http://arXiv.org/abs/0710.5420}{[arXiv:0710.5420]}.      
  
  
\bibitem{asymDM}
 D.~E.~Kaplan, M.~A.~Luty and K.~M.~Zurek,
  Phys.\ Rev.\ D {\bf 79}, 115016 (2009)
         \href{http://arXiv.org/abs/0901.4117}{[arXiv:0901.4117]}.      

\bibitem{Allahverdi:2013mza} 
  R.~Allahverdi, B.~Dutta and ,
         \href{http://arXiv.org/abs/1304.0711}{arXiv:1304.0711}.    



\bibitem{Jungman:1995df} 
  G.~Jungman, M.~Kamionkowski and K.~Griest,
  Phys.\ Rept.\  {\bf 267}, 195 (1996).

    
\bibitem{Kumar:2012uh} 
  J.~Kumar, J.~G.~Learned, S.~Smith and K.~Richardson,
  Phys.\ Rev.\ D {\bf 86}, 073002 (2012)
          	  \href{http://arXiv.org/abs/1204.5120}{[arXiv:1204.5120]}.      
  


  

\bibitem{griest}
  K.~Griest and D.~Seckel,
  Nucl.\ Phys.\ B {\bf 283}, 681 (1987)
  [Erratum-ibid.\ B {\bf 296}, 1034 (1988)].

\bibitem{solar}
  A.~Serenelli, S.~Basu, J.~W.~Ferguson and M.~Asplund,
  Astrophys.\ J.\  {\bf 705}, L123 (2009)
      	  \href{http://arXiv.org/abs/0909.2668}{[arXiv:0909.2668]}.      

\bibitem{kw}
  R.~Kappl and M.~W.~Winkler,
  Nucl.\ Phys.\ B {\bf 850}, 505 (2011)
        	  \href{http://arXiv.org/abs/1104.0679}{[arXiv:1104.0679]}.      

\bibitem{gondolo} 
  P.~Gondolo and G.~Gelmini,
  Nucl.\ Phys.\ B {\bf 360}, 145 (1991).
 
\bibitem{gould87}
   A.~Gould,
  Astrophys.\ J.\  {\bf 321}, 560 (1987).

\bibitem{Gould:1987ir}
  A.~Gould,
  Astrophys.\ J.\  {\bf 321} (1987) 571.

\bibitem{Wikstrom:2009kw}
  G.~Wikstrom and J.~Edsjo,
  JCAP {\bf 0904}, 009 (2009)
    	  \href{http://arXiv.org/abs/0903.2986}{[arXiv:0903.2986]}.      

\bibitem{eder}
G.~Eder, Nuclear Forces (MIT Press, 1968).

\bibitem{Lewin:1995rx}
  R.~H.~Helm,
  Phys.\ Rev.\  {\bf 104}, 1466 (1956);
  J.~D.~Lewin and P.~F.~Smith,
  Astropart.\ Phys.\  {\bf 6} (1996) 87.

\bibitem{Duda:2006uk}
  G.~Duda, A.~Kemper and P.~Gondolo,
  JCAP {\bf 0704} (2007) 012
      	  \href{http://arXiv.org/abs/hep-ph/0608035}{[hep-ph/0608035]}.      

\bibitem{gould90}
   A.~Gould,
  Astrophys.\ J.\  {\bf 356}, 302 (1990).
  
\bibitem{gilliland}
R.~L.~Gilliland, J.~Faulkner, W.~H.~Press, D.~N.~Spergel,
 Astrophys.\ J.\  {\bf 306}, 703 (1986);
 M.~Nauenberg,
  Phys.\ Rev.\ D {\bf 36}, 1080 (1987).  

\end{thebibliography}
\end{document}